\documentclass[aip,pof, preprint]{revtex4-1}

\usepackage{amsmath}
\usepackage{graphicx}
\usepackage{upgreek}

\newcommand{\be}{\begin{equation}}
\newcommand{\ee}{\end{equation}}

\newcommand{\bs}{\begin{subequations}}
\newcommand{\es}{\end{subequations}}

\newcommand{\ba}{\begin{array}{c}}
\newcommand{\ea}{\end{array}}

\newcommand{\bd}[1]{\mathbf{#1}}

\newcommand{\bsym}[1]{{\boldsymbol #1}}

\newcommand{\br}{\bd{r}}
\newcommand{\bk}{\bd{k}}
\newcommand{\bv}{\bd{v}}
\newcommand{\bK}{\bd{K}}

\newcommand{\efac}{\rme^{i\bk\cdot\br}}
\newcommand{\emfac}{\rme^{-i\bk\cdot\br}}

\newcommand{\rmd}{\mathrm{d}}
\newcommand{\rmD}{\mathrm{D}}
\newcommand{\rme}{\mathrm{e}}

\newcommand{\Frs}{\mathrm{Fr}_S}
\newcommand{\Bd}{\mathrm{Bd}}

\newcommand{\half}{{\textstyle \frac1{2}}}

\begin{document}

\title{Initial surface disturbance on a shear current: the Cauchy--Poisson problem with a twist}
\author{Simen \AA.\ Ellingsen} 
\affiliation{Department of Energy and Process Engineering, Norwegian University of Science and Technology, N-7491 Trondheim, Norway}

\begin{abstract}
We solve for the first time the classical linear Cauchy--Poisson problem for the time evolution an initial surface disturbance when a uniform shear current is present beneath the surface. The solution is general, including the effects of gravity, surface tension and constant finite depth. The particular case of an initially Gaussian disturbance of width $b$ is studied for different values of three system parameters: a ``shear Froude number'' $S\sqrt{b/g}$ ($S$ is the uniform vorticity), the Bond number and the depth relative to the initial perturbation width.  Different phase and group velocity in different directions yield very different wave patterns in different parameter regimes when the shear is strong, and the well known pattern of diverging ring waves in the absence of shear can take on very different qualitative behaviours. For a given shear Froude number, both finite depth and nonzero capillary effects are found to weaken the influence of the shear on the resulting wave pattern. The various patterns are analysed and explained in light of the shear-modified dispersion relation.
\end{abstract}
\maketitle

\section{Introduction}

One of the classic problems in the field of fluid mechanical surface waves goes back to Cauchy \cite{cauchy1827} and Poisson \cite{poisson1816}: A fluid surface is prescribed an initial shape and velocity whereupon it is 
left to its own devices, and the subsequent surface motion is studied. It is among the earliest endeavors in water wave theory, and the independent treatments by Cauchy and Poisson paved the way for much of the subsequent progress\cite{craik04}. The solutions of Cauchy and Poisson apprared at a time where the mathematics of waves was rapidly progressing, and is very well recounted in Darrigol's fascinating historical review\cite{darrigol03}.

The most well known manifestation of the Cauchy--Poisson problem is probably a localised initial disturbance, such as may be produced by throwing a pebble into a pond, and the resulting pattern of ring-shaped waves is well known to all. By introducing boundaries and/or non-uniform depth, far more involved initial value problems may be studied as indeed they have been over the two centuries which have passed since the initial works\cite{thomson1887,finkelstein57,miles68,rhodes-robinson84,debnath89}. Excellent reviews covering most of the classical literature up to the time of their publication are Lamb's classical text\cite{lamb32} and the famous review of Wehausen and Laitone \cite{wehausen60}. Of practical applications of the Cauchy--Poisson initial value problem, the most striking is that of tsunamis \cite{voit87}, giant destructive waves developing from an initial impulse imparted by an underwater earthquake, and similarly underwater explosions \cite{kranzer59}.

\begin{figure}[htb]
  \includegraphics[width=2.7in]{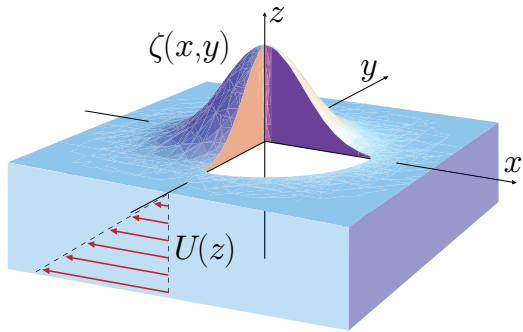}
  \caption{The geometry considered: a small initial disturbance (here in the shape of a Gaussian bell) is left to oscillate freely into ring waves on top of a velocity profile of constant shear.}
  \label{fig_geom}
\end{figure}

The mathematical framework developed for Cauchy--Poisson problems in terms of Green's functions is quite general and powerful, as laid out in section 22 of Ref.~\onlinecite{wehausen60}. This theory does, however, rely on the use of a velocity potential, which in 3 dimensions limits the theory to situations of irrotational flow \cite{ellingsen13a}. It was recently realised, however, that the Euler equations permit an exact solution for linear surface waves in the presence of uniform shear, allowing the study of ship waves in the presence of uniform vorticity \cite{ellingsen14}. For that case the presence of a shear current was found to be able to change a ship's train of waves quite radically.

The dispersion relation in the presence of uniform shear reported in Ref.~\onlinecite{ellingsen14} shows that group and phase velocities can differ greatly in different directions of propagation when the uniform shear (in suitably non-dimensionalised form) is strong. Thus one must expect that the well known circular ring waves, such as seen when throwing a pebble into a quiescent pond, will no longer be circular and perhaps not even ring shaped, with a shear current present. 

While three dimensional linear wave theory with uniform shear seems to be a recent progression, much work has been performed in two dimensions on waves superposed on a Couette profile. In a classical paper, Taylor considered such a model to investigate the wave stopping power of a sheet of rising bubbles \cite{taylor55}, an application followed up in later works \cite{brevik79,brevik93}. A particular branch of research has focused on the somewhat exotic analogy between such wave stopping and event horizons near black holes \cite{nardin09,rousseaux13}. The literature on interactions between waves and a uniform--vorticity current in two dimensions is significant, and a selection is found in Refs.~\onlinecite{ellingsen13a, taylor55,brevik79,brevik93,biesel50,thompson50,fenton73,peregrine76,johnsson78,telesdasilva88,fabrikant98,vanden-broeck00, margaretha05,benzaquen12,tyvand14}. 

In the present work we lay out the linearised theory of a general initial disturbance (as given by a prescribed surface shape and velocity at $t=0$), and thereupon conduct a numerical study of the patterns resulting from an initially static and symmetrical Gaussian perturbation in different parameter regimes, investigating the effect of finite depth, surface tension vs gravity driven waves (as described by the Bond number) and the strength of the vorticity.

\section{General formalism and dispersion relation}

We consider the geometry depicted in figure \ref{fig_geom}. A surface which, when undisturbed, is at $z=0$, is disturbed at $t=0$ so that its surface elevation at that instance is described by $\zeta(\br,0)$ and $\dot\zeta(\br,0)$ where the latter is the surface motion at $t=0$. We work to linear order in the surface elevation, which is assumed small. Here $\br = (x,y)$ is the position in the horizontal plane. Beneath the surface there is a linear (Couette) shear flow along the $x$ axis with constant vorticity $\bsym{\omega}=-S\hat{\bd{y}}$ (a hat denotes a unit vector),
\be\label{U}
  U(z) = U_0 + S z.
\ee
The water is assumed to have depth $h$. The flow is not necessarily viewed from the intertial system wherein the bottom is at rest, hence there are no particular restrictions on the relative magnitudes of $U_0, S$ and $h$. We will assume $S\geq 0$ for definiteness. We shall assume the liquid to be homogeneous and incompressible, and the effects of viscosity are neglected.

We proceed by combining the methods of Refs.~\onlinecite{ellingsen13a} and \onlinecite{ellingsen14}. The full velocity field we write
\be
  \bv(\br,z,t) = (U(z) + \tilde{u}, \tilde{v}, \tilde{w}); ~~~ p(\br,z,t) = -\rho g z + \tilde{p}
\ee
where $\tilde{u},\tilde{v},\tilde{w}$ and $\tilde{p}$ are the perturbations stemming from the initial deformation.
We take the Fourier transform in the $xy$ plane and assume plane wave form for the small perturbations so that the full physical solution can be written on the form
\be
  \ba\tilde{u}(x,y,z,t)\\\tilde{v}(x,y,z,t)\\\tilde{w}(x,y,z,t)\\\tilde{p}(x,y,z,t)\ea = \int\frac{\rmd^2k}{(2\pi)^2}\left[\ba u(\bk,z,t)\\v(\bk,z,t)\\w(\bk,z,t)\\p(\bk,z,t)\ea\right]\efac
\ee
where 
\[
  \bk=(k_x,k_y) = (k\cos\theta_k,k\sin\theta_k)
\]
is the 2-dimensional wave vector. We must bear in mind that $u,v,w,p$ are now in general complex quantities while $\tilde u,\tilde v,\tilde w,\tilde p$ are of course real. Inserting into the Euler equations, 
\[
  \frac{\rmD\bv}{\rmD t} = -\frac1{\rho}\nabla p - g \hat{\mathbf{z}}
\]
($\rmD/\rmD t$ is the material derivative) to linear order gives
\bs
\begin{align}
  \dot{u} + i k_x U(z) u + S w =& -i k_x p/\rho \label{Ea}\\
  \dot{v} + i k_x U(z) v =& -i k_y p/\rho \label{Eb}\\
  \dot{w} + i k_x U(z) w =& -p'/\rho \label{Ec}\\
  ik_x u+ik_yv+w' =&0\label{Ed}
\end{align}
\es
where a dot denotes a time derivative and a prime a derivative with respect to $z$.

We now carry out the tedious process of eliminating $p,u$ and $v$ [differentiating \eqref{Ea} and \eqref{Eb} w.r.t.\ $z$ eliminates $p$ and one surface-parallel velocity component, then the other component is eliminated using \eqref{Ec}], ending up with a Rayleigh equation on the form
\be
  [\partial_t + i k_xU(z)](\partial_z^2-k^2)w = 0.
\ee
This implies
\be\label{inhom}
   (\partial_z^2-k^2)w = k^3D(\bk) \rme^{-i k_xU(z)t}
\ee
[$D$ is an arbitrary function; coefficients are multiplied by powers of $k$ for dimensional reasons] whose general solution we may choose to write
\[
  w(z,t) = kA(\bk,t)\sinh k(z+h) + kC(\bk,t)\cosh k(z+h) + \frac{k^3D(\bk)\rme^{-i k_xU(z)t}}{k^2+k^2_xS^2 t^2}.
\]
Here $A,C$ are undetermined coefficients of the homogeneous solution, and the last term is the particular solution of \eqref{inhom}. Neither the term $\propto \cosh k(z+h)$ nor the particular solution fulfil the boundary condition $w(-h,t)=0$, hence $C=D=0$. Thus we are left with 
\be
  w(z,t) = kA(\bk,t)\sinh k(z+h).
\ee
Inserting back via \eqref{Ec}, we may write the pressure disturbance as
\be\label{peq}
  p(z,t)/\rho = -(\dot{A} + i k_xUA)\cosh k(z+h)+i S A \cos \theta_k \sinh k(z+h)+\text{const.}
\ee

We define the surface elevation in the same manner, in terms of its Fourier transform,
\be\label{zetadef}
  \zeta(\br,t) = \int\frac{\rmd^2k}{(2\pi)^2}B(\bk,t)\efac. 
\ee
Since $\zeta$ is a real quantity it follows that
\begin{align*}
  \int\frac{\rmd^2k}{(2\pi)^2}B(\bk,t)\efac=&\frac12\int\frac{\rmd^2k}{(2\pi)^2}[B(\bk,t)\efac+B^*(\bk,t)\emfac]\\
  =&\frac12\int\frac{\rmd^2k}{(2\pi)^2}[B(\bk,t)+B^*(-\bk,t)]\efac
\end{align*}
($^*$ denotes complex conjugate) so $B$, like any Fourier transform of a real quantity, satisfies the symmetry
\be\label{symm}
  B(-\bk,t) = B^*(\bk,t).
\ee

The linearised kinematic boundary condition at the surface is $\mathrm{D}\zeta/\mathrm{D}t =(\partial_t+ik_xU_0)\zeta = w(0,t)$, i.e.,
\be\label{kin}
  \dot{B}+ik_xU_0B = kA\sinh kh.
\ee
The dynamic boundary condition at the surface is that the pressure just \emph{above} the interface equals the atmostpheric pressure, which we set to zero. Including, and linearising, the pressure jump from surface tension, $\Delta p_\text{s.\,t.}=\sigma(\partial_x^2+\partial_y^2)\zeta$ ($\sigma$ is the surface tension coefficient). In Fourier space, $\sigma(\partial_x^2+\partial_y^2)\zeta\to-\sigma k^2 B$, and inserting the expression \eqref{peq} for $p$ just below the interface, the boundary condition reads
\be
  (\dot{A}+ i k_x U_0 A) \cosh kh - i S A \cos\theta_k\sinh kh +(g+\sigma k^2/\rho)B=0.
\ee

Eliminating $A$ by means of Eq.~\eqref{kin} we obtain the time evolution equation for the surface elevation, 
\be\label{beq}
  \ddot{B}+2i \omega_1 \dot{B}+\omega_2^2B=0,
\ee
where we have defined
\bs
\begin{align}
  \omega_1=&k_xU_0-\half S \tanh kh \cos\theta_k; \\
  \omega_2^2=& (k^2c_0^2+kSU_0\cos^2\theta_k)\tanh kh - k_x^2 U_0^2;\\
  c_0^2 =& \left(\frac{g}{k}+\frac{k\sigma}{\rho}\right)\tanh kh.
\end{align}
\es
Here $c_0$ is the well known expression for the phase velocity in the absence of shear (c.f.\ e.g.\ Ref.~\onlinecite{wehausen60}). The solutions to Eq.~\eqref{beq} are thus $\propto \exp(-i\omega_1t\pm i\sqrt{\omega_1^2+\omega_2^2}t)$, which we write
\be\label{Bprop}
  B \propto \rme^{-i(k_xU_0+\omega_\pm)t}
\ee
where
\be
  \omega_{\pm} = -\half S \tanh kh \cos\theta_k\pm\sqrt{c_0^2k^2+(\half S\tanh kh \cos\theta_k)^2}.
\ee

We have excluded the uniform flow term $k_xU_0$ from the definition of $\omega_\pm$ because this term is physically trivial in the present sense, and can always be transformed out of the problem by simply going to the coordinate system $\br \to \br - U_0t\hat{\bf{x}}$ where the surface is at rest. Physically this means that the wave pattern once initiated is convected passively with the flow at velocity $U_o\hat{\bf{x}}$. The situation is thus radically different than that of a continuously emitting source, say an oscillating source, in which case Doppler frequency shifts would be present. That problem, important for marine hydrodynamic purposes, we shall consider in the near future.

\subsubsection{More general velocity profiles}

The above analysis can be performed exactly courtesy of the linear velocity profile \eqref{U}. For a more general $U(z)$ the situation is much complicated by the formation of critical layers; in two dimensions the vertical velocity perturbation $w$ is known to satisfy the Rayleigh equation (e.g.\ Ref.\ \onlinecite{leblond78} Ch.\ 7)
\be\label{rayleigh}
  w'' -\left[\frac{U''(z)}{U(z)-c(k)}+k^2\right]w=0
\ee
which generalises the relation $w''-k^2w=0$ which was found above. At a depth $z_c$ so that $U(z_c)=c(k)$ the Rayleigh equation is singular if $U''(z_c)\neq 0$, and a so--called critical layer results, where cat's eye vortices form\cite{leblond78} and the linearized and inviscid solution becomes inaccurate. A critical layer attenuates waves passing through it, reflecting wave energy back towards the surface\cite{booker67}. A corresponding effect in the boundary layer formed by wind above a water surface is held to be an important factor in the generation of waves by wind\cite{miles57,hristov02}.

Typical velocity profiles to encounter in real flows include parabolic and Blasius--type profiles, both of which permit the formation of critical layers. It is not obvious what the effect might be, since an initial disturbance corresponds to a distribution of plane waves of different wavelengths, each of whose critical layers (if they exist) will be found at different depths. One might hazard a conjecture, however, that the effect of critical layers will be important in cases where the most important wavelengths of the disturbance result in critical layers near the surface, whereas in cases where the critical depth $z_c$ for the most important wave number $k_0$ is either deep ($|k_0z_c|\gg 1$) or does not exist ($U(z)$ never equals $c(k_0)$) the linear theory found herein is a reasonable approximation of the physical picture. This important question will, however, require much futher investigation.

\subsection{Dispersion relation}

For the present purposes it is natural to define the phase velocity (and later, group velocity) relative to the surface of the water, which is why the term $-i k_xU_0$ is excluded from from the definition of $\omega_\pm$ in Eq.~\eqref{Bprop}. We see that for each wave vector $\bk$ we have one solution of frequency $\omega_+(\mathbf{k})>0$ propagating in direction $\mathbf{k}$ (positive phase velocity), and one solution of frequency $\omega_-(\mathbf{k})<0$ which corresponds to a wave  propagating with negative phase velocity $\omega_-(\mathbf{k})/k$ along direction $\bk$. In other words the latter solution describes a wave moving at phase velocity $|\omega_-(\mathbf{k})|/k=-\omega_-(\mathbf{k})/k$ in direction $-\mathbf{k}$. However, we note that $-\omega_-(\mathbf{k})=\omega_+(-\mathbf{k})$, meaning that there is only one possible phase velocity for a wave moving in a given direction of propagation $\mathbf{k}$, and the dispersion relation is
\be  \label{c}
  c(\mathbf{k}) = \sqrt{c_0^2+\Bigl(\frac{S}{2k}\tanh kh\cos\theta_k\Bigr)^2}-\frac{S}{2k}\tanh kh\cos\theta_k
\ee
as also given in Ref.~\onlinecite{ellingsen14}. This dispersion relation generalises the dispersion relation in the 2D geometry found in the literature, see, e.g., Ref.~\onlinecite{ellingsen13a,benzaquen12}, which is regained when $\theta_k=0,\pi$. For an in--depth discussion of the dispersion relation for gravity waves in the 2D geometry, c.f.\ Ref.\ \onlinecite{margaretha05}.

The group velocity in direction $\hat{\bk}$ is now given as
\be
  c_g(\mathbf{k}) = (\hat{\bf{k}}\cdot\nabla_k) kc(\mathbf{k})=\frac{\rmd }{\rmd k} kc(\mathbf{k}),
\ee
which is straightforward to calculate but not reproduced explicitly in the most general case due to the bulkiness of the resulting expression. We will, however, give explicit group velocity expressions in various limiting cases in the following.

\subsection{General solution to initial value problem}

We are now ready to solve the problem of an initial surface disturbance. By summing over all possible values of the wave vector $\bf{k}$, we can write the surface deformation in the general form using Eq.~\eqref{Bprop},
\be\label{zetagen}
  \zeta(\br,t) = \int\frac{\rmd^2 k}{(2\pi)^2}e^{i\bk\cdot\br-ik_xU_0t}\left[\beta_+(\bk)\rme^{-i\omega_+t}+\beta_-(\bk)\rme^{-i\omega_-t}\right]
\ee
where $\beta_{\pm}$ are undetermined complex-valued coefficients.
Consequently,
\begin{align}
  \dot{\zeta}(\br,t) =& -i\int\frac{\rmd^2 k}{(2\pi)^2}e^{i\bk\cdot\br-ik_xU_0t}\bigl[(k_xU_0+\omega_+)\beta_+(\bk)\rme^{-i\omega_+t}\notag \\
  &+(k_xU_0+\omega_-)\beta_-(\bk)\rme^{-i\omega_-t}\bigr].\label{zetadotgen}
\end{align}


In the initial value problem, $\zeta(\br,0)$ and $\dot{\zeta}(\br,0)$ are assumed to be given quantities. The Fourier transforms of the initial deformations are then also known quantities which we define:
\be\label{init}
  B_0(\bk)=\int\rmd^2 r \zeta(\br,0)\emfac ; ~~~
  \dot{B}_0(\bk)=\int\rmd^2 r \dot{\zeta}(\br,0)\emfac.
\ee
[This matches the previous definition, Eq.~\eqref{zetadef}].
Hence Eqs.~\eqref{zetagen} and \eqref{zetadotgen}, and the linear independence of different Fourier components imply
\begin{align*}
  \beta_+(\bk)+\beta_-(\bk)=& B_0(\bk); \\
  (k_xU_0+\omega_+)\beta_+(\bk)+(k_xU_0+\omega_-)\beta_-(\bk)=& i \dot{B}_0(\bk).
\end{align*}
Solving these with respect to the $\beta$ coefficients gives the solution to the general initial value problem:
\be
 \beta_\pm(\bk) = \frac{\mp1}{\omega_+-\omega_-} [(k_xU_0+\omega_\mp)B_0(\bk)-i\dot B_0(\bk)].
\ee
So far the only restriction is to assume that the Fourier transforms $B_0$ and $\dot B_0$ exist.

A somewhat more physically transparent formulation results if we exercise our freedom to rotate the integrand by an angle $\pi$ in the $\bk$ plane by letting $\bk\to -\bk$, using the symmetry $\omega_-(\bk) = -\omega_+(-\bk)$ and that $c(\bk) = \omega_+(\bk)/k$. Then Eq.~\eqref{zetagen} can be written
\bs
\begin{align}
    \zeta(\br,t) =& \int\frac{\rmd^2 k}{(2\pi)^2}\left[\beta_+(\bk)\rme^{i\xi(\bk,t)}+\beta_-(-\bk)\rme^{-i\xi(\bk,t)}\right];\\
   \xi(\bk,t) =& \bk\cdot\br - k_xU_0t-kc(\bk)t.
\end{align}
\es

Let us also define the velocity 
\be
  c_\text{div}(\bk) = \frac12 [c(\bk)+c(-\bk)] = \sqrt{c_0^2 + \Bigl(\frac{S}{2k}\tanh kh\cos\theta_k\Bigr)^2}
\ee
which is the velocity at which two wave crests of the same wavelength but opposite directions of propagation diverge away from the point midway between the two. 
Using the symmetry \eqref{symm} for $B_0$ and $\dot B_0$ we obtain
\begin{align}
  \zeta(\br,t) =& \int\frac{\rmd^2 k}{(2\pi)^2}\frac1{kc_\text{div}(\bk)}\Bigl\{[kc(-\bk)-k_xU_0]\mathrm{Re}\{B_0(\bk)e^{i\xi(\bk,t)}\}\notag \\
  &-\mathrm{Im}\{\dot B_0(\bk)e^{i\xi(\bk,t)}\}\Bigr\}.
\end{align}

In case of an initial disturbance which is symmetrical under point inversion we have $B_0(-\bk)=B_0(\bk)=B_0^*(\bk)$ and similarly $\dot B_0(\bk)=\dot B_0^*(\bk)$, so the $B$ coefficients are real and the expression simplifies somewhat:
\begin{align}
  \zeta(\br,t) =& \int\frac{\rmd^2 k}{(2\pi)^2}\frac1{kc_\text{div}(\bk)}\Bigl\{[kc(-\bk)-k_xU_0]B_0(\bk)\cos\xi(\bk,t) -\dot B_0(\bk)\sin\xi(\bk,t)\Bigr\}.\label{symmzeta}
\end{align}

The standard expression in the absence of shear is obtained upon setting $S=0$ and $U_0=0$ and assuming an axisymmetric initial disturbance, $\zeta(\br,0) =\zeta(r,0)$, which implies $B_0(\bk)=B_0(k)$, in which case one obtains
\bs
\begin{align}
  \zeta(r,t) =& \frac{1}{2\pi}\int_0^\infty \rmd k k J_0(kr)\left[B_0(k)\cos(kc_0 t)-\frac{\dot B_0(k)}{kc_0}\sin(kc_0t)\right],\\
  B_0(k)=& 2\pi \int_0^\infty \rmd r r \zeta(r,0)J_0(kr);~~~  \dot B_0(k)= 2\pi \int_0^\infty \rmd r r \dot\zeta(r,0)J_0(kr)
\end{align}
\es
wherein $J_0$ is the zeroth order Bessel function of the first kind.

\subsection{Gaussian initial disturbance}

In the following we shall consider the deformation from an intertial system following the surface of the water, i.e., wherein $U_0=0$. 
As an example we consider in the following the time evolution of an initially axisymmetric Gaussian surface perturbation. For ease of comparison we make use of the same form of the Gaussian as considered in the ship wave context in Refs.~\onlinecite{ellingsen14} and \onlinecite{darmon13},
\be
  \zeta(r,0) = z_0 \rme^{-\pi^2 r^2/b^2}; ~~ B_0(k) = \frac{b^2z_0}{\pi}\rme^{-k^2b^2/(2\pi)^2}.
\ee
Let us furthermore assume the perturbation initially still with respect to the water surface, $\dot\zeta(\br,0)=0$. Now Eq.~\eqref{symmzeta} gives
\be
  \zeta(\br,t) = \frac{1}{(2\pi)^2}\int_0^\infty\rmd k kB_0(k)\int_0^{2\pi}\rmd \theta_k\frac{c(-\bk)}{c_\text{div}(\bk)}\cos[\bk\cdot\br-kc(\bk)t].
\ee

It is useful to set this expression in non-dimensional form by rescaling with respect to length $b$, time $\sqrt{b/g}$ and velocity $\sqrt{bg}$. This gives the three non-dimensional system parameters
\be
   \Frs = S\sqrt{b/g}; ~~~ \Bd = b^2\rho g/\sigma;  ~~~ H = h/b 
\ee
where $\Frs$ is the ``shear Froude number'' based on the velocity $bS$,
and $\Bd$ is known as the Bond number. Also we define non-dimensional wave number and coordinates
\be
   \mathbf{K} = b\mathbf{k};~~~K=bk; ~~~ R = r/b; ~~~T = t\sqrt{g/b}; ~~~C(\mathbf{K}) = c(\mathbf{k})/\sqrt{bg}. 
\ee
With this, 
\bs\label{cdimless}
\begin{align}
  C(\mathbf{K})=&C_\text{div}(\mathbf{K}) -\frac{\Frs}{2K}\tanh KH\cos\theta_k,\\
  C_\text{div}(\mathbf{K})=&\Bigl[\Bigl(\frac1K + \frac K{\Bd}\Bigr)\tanh KH + \Bigl(\frac{\Frs}{2K}\tanh KH\cos\theta_k\Bigr)^2\Bigr]^\frac12,
\end{align}
\es
and
\begin{align}
  \frac{\zeta(\mathbf{R},T)}{z_0}=&\frac{1}{4\pi^3}\int_0^\infty\rmd K Ke^{-(K/2\pi)^2}\int_0^{2\pi}\rmd \theta_k\frac{C(-\bK)}{C_\text{div}(\bK)}\notag \\
  &\times\cos\bigr[KR\cos(\theta_k-\phi)-KC(K,\theta_k)T\bigl].
\end{align}
where we let $\mathbf{R}=(X,Y)=(R\cos\phi,R\sin\phi)$.

Note finally that with the initial perturbation $\zeta(r,0)\sim \exp[-(K/2\pi)^2]$, the main contribution to the wave pattern wil come from waves with $K\sim 2\pi$, that is, which have dimensionless wavelength $\sim 1$. When analysing dispersion in various limiting cases in the following we shall therefore consider wavelengths near unity for definiteness.

\subsection{Downstream motion of the ring pattern}\label{sec_circs}

\begin{figure}[htb]
  \includegraphics[width=.5\textwidth]{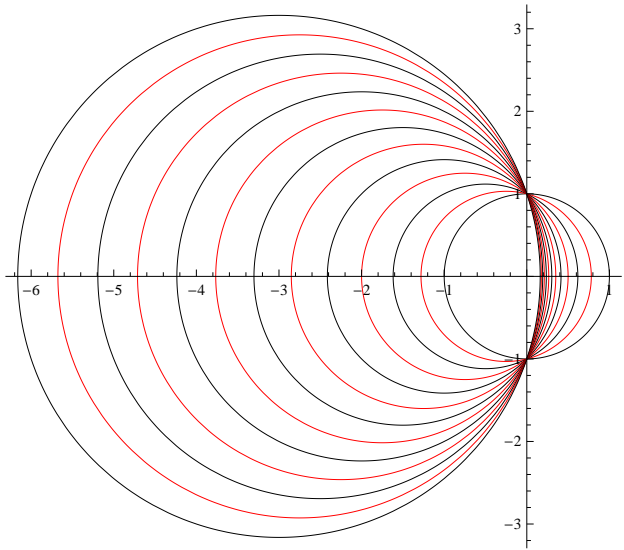}
  \caption{Polar plots of $c(k,\theta)/c_0(k)$ form circles centered a distance $a$ downstream, of radius $\sqrt{1+a^2}$. Here $a$ [see Eq.~\eqref{a}] ranges from $0$ (smallest circle) to $3$ in steps of $0.25$.}
  \label{fig_circles}
\end{figure}

Consider the shape drawn by a single wave crest moving at phase velocity. At a time $t$ it draws a curve $r(\theta) = c(k,\theta)t$. Now note that $c$ is of the form 
\be
  c(k,\theta) = c_0(k)\Bigl[\sqrt{1+a^2\cos^2\theta}-a\cos\theta\Bigr]
\ee
where
\be\label{a}
  a = \frac{S}{2}\sqrt{\frac{\tanh kh}{k(g+k^2\sigma/\rho)}}.
\ee
Defining coordinates $\xi = r(\theta)\cos\theta + ac_0t$ and $\eta = r(\theta)\sin\theta$, one easily ascertains that  $\xi^2+\eta^2=c_0^2t^2(1+a^2)$, so the curve drawn by a crest moving outwards at phase velocity is a circle of radius $c_0t\sqrt{1+a^2}$ centered at $(x,y)=(-ac_0t,0)$. The radius of each circle grows relative to its moving centre at a velocity $c_\text{radius}=c_0$.

\begin{figure}[htb]
  \includegraphics[width=\textwidth]{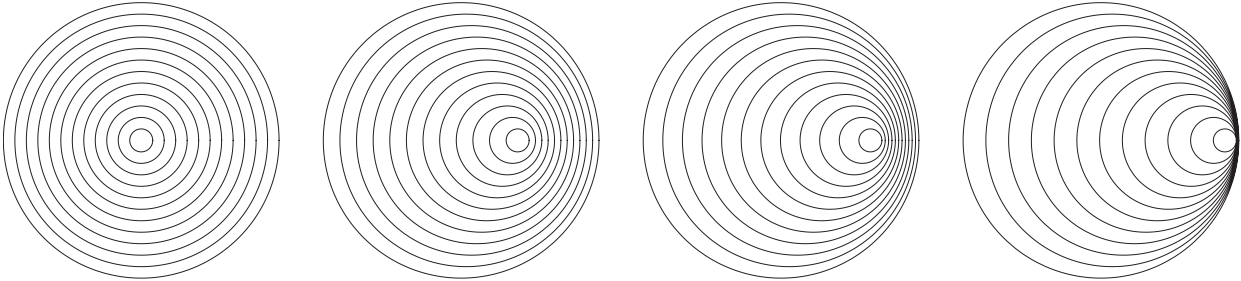}
  \caption{Curves $r(\theta)=c(k,\theta)t$ at a series of equidistant times. Left to right: $a=0,0.5,1,5$.}
  \label{fig_circlemove}
\end{figure}

One effect of the presence of the shear current is therefore to simply shift the ring pattern at wavenumber $k$ downstream at a constant rate $c_0(k)a t$ as shown in Fig.~\ref{fig_circlemove}. This picture is unable to fully explain the various phenomena observed in the following, however, which require a more careful consideration of how the shear affects dispersion. The interplay between phase and group velocities in different directions will be seen to be of particular importance.

\section{Numerical examples and limiting cases}

In the following we explore the transient evolution of a Gaussian initial perturbation in various parameter regimes. For analysis we must turn time and again to the dispersion relation \eqref{cdimless}.
The dimensionless group velocity for a plane wave propagating in direction $\bK/K$ is
\be\label{Cg}
  C_g(\mathbf{K}) = \frac{\rmd}{\rmd K}[KC(\mathbf{K})].
\ee
The general expression for $C_g$ is bulky, but becomes more handy in various far corners of our parameter space.

\begin{figure}[htb]
  \includegraphics[width=\textwidth]{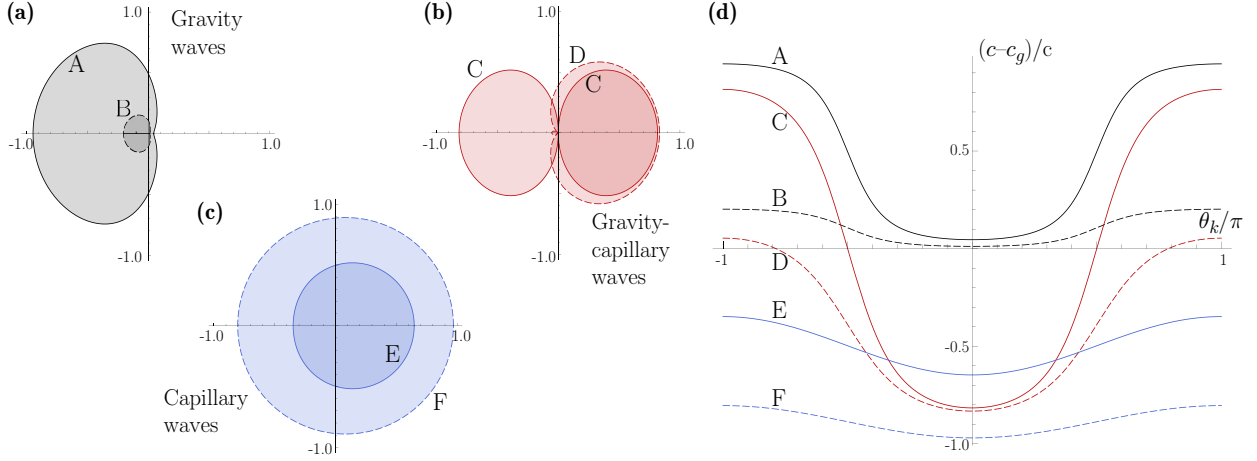} 
  \caption{(a,b,c) Polar plots of $|c-c_g|/c$ as a function of $\theta_k$, and (d) plot of $(c-c_g)/c$, for different parameter combinations considered in the following sections. In all graphs, $\Frs=10$ and velocities are plotted at $k=2\pi b$. Gravity waves ($\Bd=\infty$) at deep water $H=10$ (A) and shallow water $H=0.1$ (B), capillary--gravity waves ($\Bd=4\pi^2$) at deep water $H=10$ (C) and shallow water $H=0.2$ (D), and capillary waves ($\Bd=0.1$) at deep water $H=10$ (E) and shallow water $H=0.1$ (F).}
  \label{fig_cdif}
\end{figure}

For all the different parameter combinations considered in the following we plot the relative difference between phase and group velocity, shown in Fig.~\ref{fig_cdif}. Discussion of the different curves in this figure are found in the respective sections treating each parameter regime.

\begin{figure}[htb]
  \includegraphics[width=\textwidth]{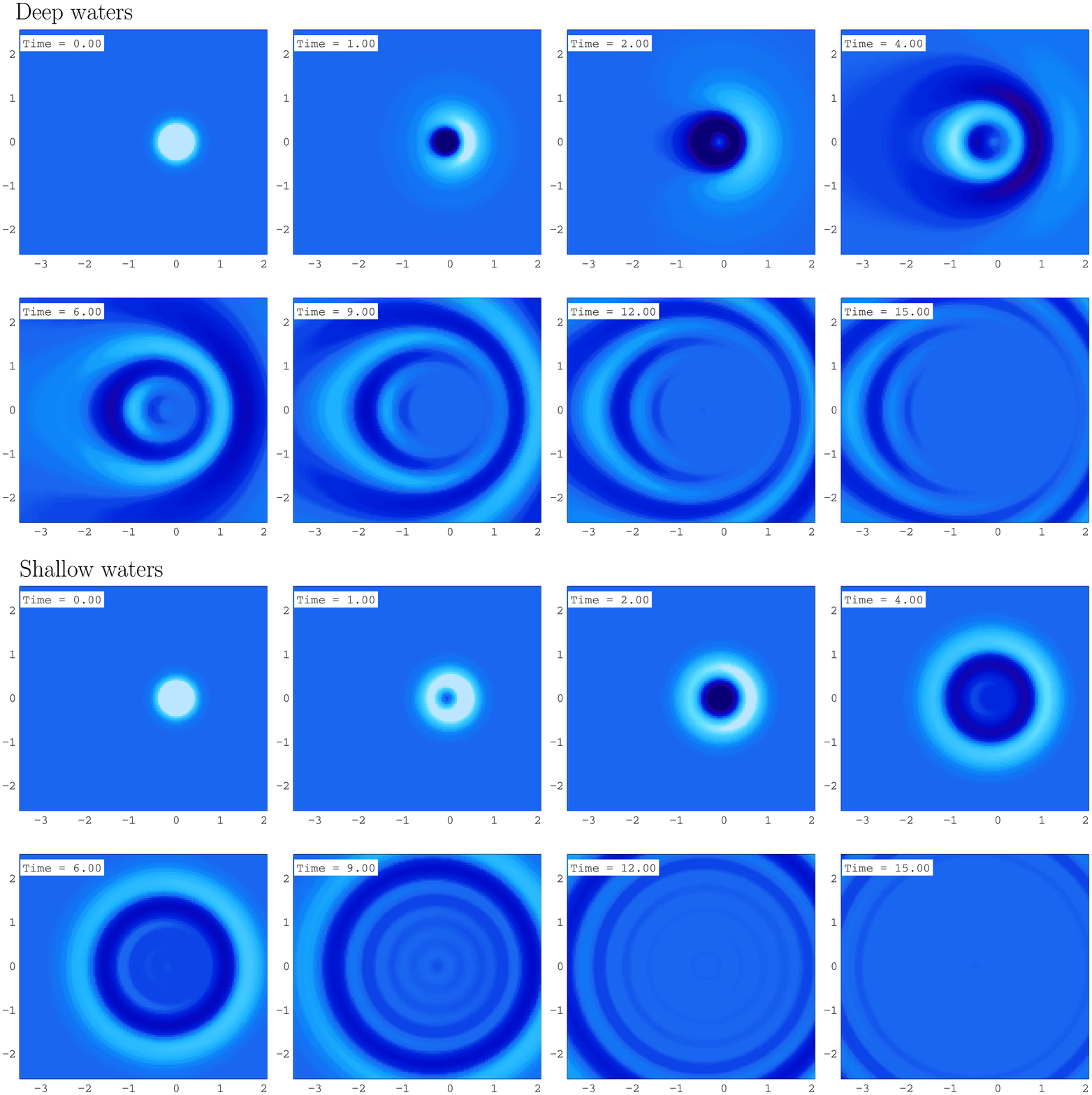}
  \caption{The effect of water depth at moderate shear for gravity waves shown at different times $T$. In all panels $\Frs=1$ and $\Bd=\infty$. Top 8 panels: $H=10$, bottom 8 panels: $H=0.1$. For improved contrast, areas in this and all below figures where $\zeta>0.2z_0$ are white, $\zeta<-0.2z_0$ are black, with linear colour gradient for amplitudes in between. (Multimedia view)}
  \label{fig_moderate}
\end{figure}

\subsection{The effect of depth for pure gravity waves}

We begin by investigating the effect of depth for pure gravity waves, i.e., $\Bd=\infty$. When the water is deep, the dispersion relation then tends to
\be
  C(\mathbf{K})\buildrel{H\to\infty}\over{\longrightarrow}\Bigl[\frac1K + \Bigl(\frac{\Frs}{2K}\cos\theta_k\Bigr)^2\Bigr]^\frac12 -\frac{\Frs}{2K}\cos\theta_k,
\ee
while for water shallow compared to disturbance width, $H\ll1$, the dispersion relation is independent of $K$:
\be
  C(\mathbf{K})\buildrel{H\ll1}\over{\longrightarrow}\Bigl[H + \Bigl(\frac{\Frs H}{2}\cos\theta_k\Bigr)^2\Bigr]^\frac12 -\frac{\Frs H}{2}\cos\theta_k.
\ee
Corresponding limiting expressions for the group velocity are 
\be
  C_g(\mathbf{K})\buildrel{H\to\infty}\over{\longrightarrow} \frac{1}{2\sqrt{K+(\Frs\cos\theta_k/2)^2}} 
\ee
and
\be
  C_g(\mathbf{K})\buildrel{H\ll1}\over{\longrightarrow} C(\mathbf{K}).
\ee

Inspection reveals that in deep waters the phase velocity exceeds the group velocity in all directions for all $K>0$, but the difference between the two will depend on the direction of propagation. When the shear is strong enough the waves will therefore not stay ring shaped, but break into crescents as the crests and troughs have different life span in different propagation directions. In shallow water, on the other hand, the phase and group velocities tend to the same value and the wave propagating outwards will be relatively stable crests which remain ring-shaped (though not in general circular) and long-lived. 

The situation is shown in figure \ref{fig_moderate} where a moderate shear, $\Frs=1$, is considered for gravity waves in deep ($H=10$) and shallow ($H=0.1$) waters. The qualitative behaviour derived from considering the dispersion relation are clear to see; effects of the shear current are more prominent in deep waters.

\subsubsection{Strong shear, deep water}

In order to study the effect of shear, we consider in the following a large value of the shear Froude number, $\Frs=10$. While such strong shear near the surface may not be straightforward to create in experiment, it serves our purposes here by exaggerating the effect of the shear on the ring wave pattern.

We first consider the case of deep water and no surface tension and assume $\Frs\gg1$. The shear $S$ enters the dispersion relation only through the combination $\Frs\cos\theta_k$, so in directions close to $\theta_k=\pm\pi/2$, the shear current has little effect upon the wave velocities. Consider therefore directions where $|\cos\theta_k|\gg 1/\Frs$. Keeping leading orders in $1/\Frs$ we have
\begin{subequations}\label{dispgravdeep}
\begin{align}
  C(\mathbf{K}) =& \frac{\Frs}{2K}(|\cos\theta_k|-\cos\theta_k) + \frac{1}{\Frs|\cos\theta_k|} + ... \\
  C_g(\mathbf{K}) =& \frac{1}{\Frs|\cos\theta_k|} +...
\end{align}
\end{subequations}
Thus, in directions $\cos\theta_k<0$ we have $C\gg C_g$, differing gratly from directions $\cos\theta_k>0$ where instead $C\approx C_g$ up to a difference of order $\Frs^{-2}$. 

\begin{figure}[htb]
  \includegraphics[width=\textwidth]{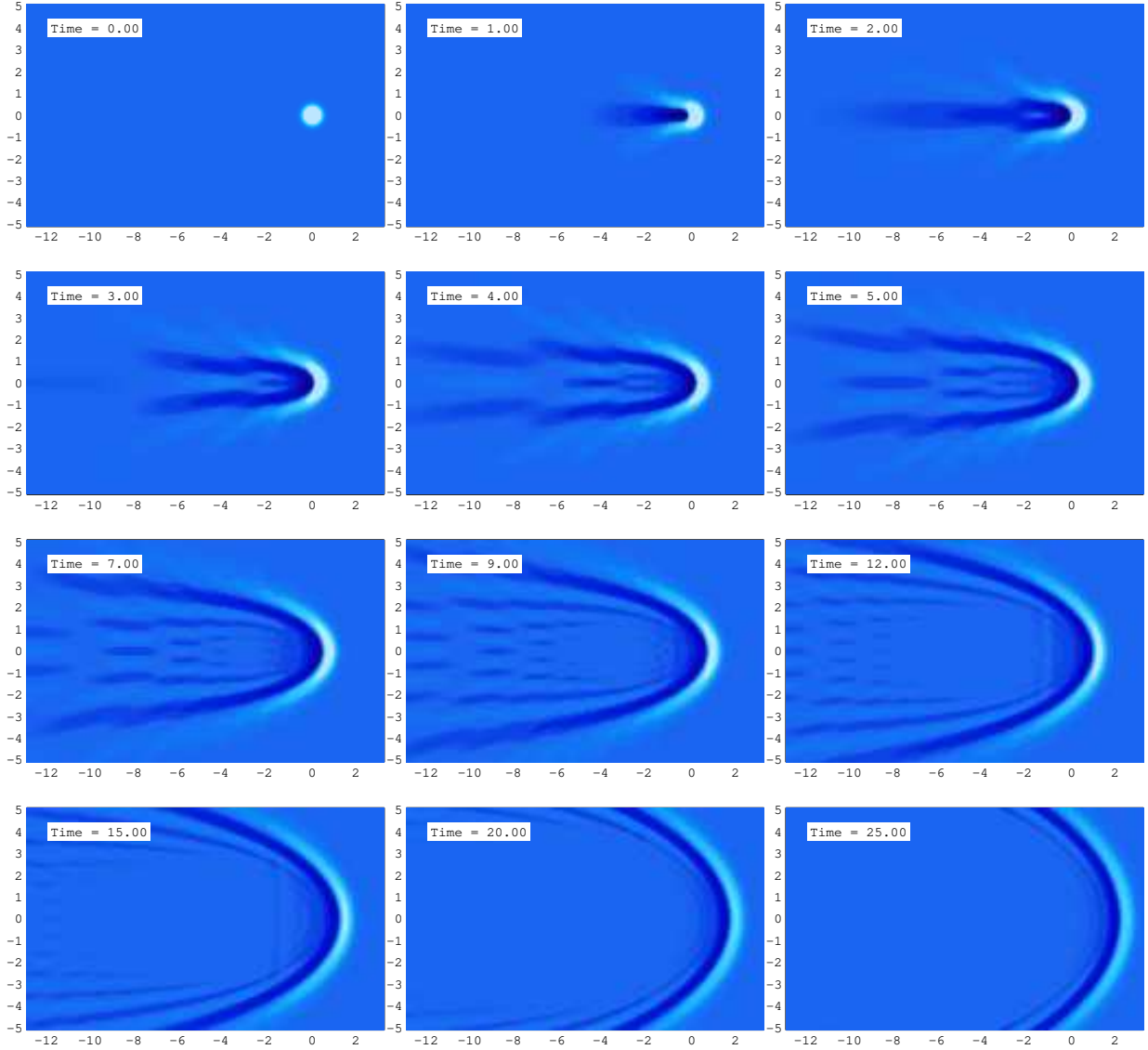}
  \caption{Relief plots of gravity waves with strong shear in deep waters shown at different times $T$, parameters $\Frs=10$, $\Bd=\infty$, $H=10$. Colour shading as in Fig.~\ref{fig_moderate}. (Multimedia view)}
  \label{fig_gravDeep}
\end{figure}

The behaviour of the wave pattern is now very different upstream vs downstream, as made clear from a relief plot at different times $T$, shown in Fig.~\ref{fig_gravDeep}. The shear flow beneath the surface is from right to left. Waves moving against the current towards the right ($\cos\theta_k>0$) have phase velocity $C\approx C_g \ll 1$, so there is a single, long-lived wave crest, greatly slowed down by the current. Downstream, however, the group velocity is similarly slow ($\sim \Frs^{-1}$), but the phase velocity is much greater ($\sim\Frs$), as also seen in figure \ref{fig_cdif} (graph A). The result is a flapping appearance with crests and troughs in rapid motion, originating near the disturbance and quickly disappearing again downstream. 

\subsubsection{Shallow water, moderate shear}

Consider now what happens in shallow water, i.e., $H\ll 1$. Again we let $\Bd=\infty$ (pure gravity waves). For (dimensionless) wavelengths of order $K\sim 2\pi$ (wavelengths similar to size of initial disturbance) which are the most important we may Taylor expand: $\tanh KH= KH-(KH)^3/3+...$. When the shear is \emph{moderate}, $\Frs \sim 1$ or less, the phase velocity becomes,
\begin{align}
  C(\mathbf{K})\sim&\sqrt{H}-\frac{\Frs H}{2}\cos\theta_k + ... 
\end{align}
Since $C$ is largely independent of $K$, it follows from \eqref{Cg} that $C_g\approx C$.
More specifically, the phase velocity is independent of $K$ up to order $H^{5/2}$, and hence 
\be
  C(\mathbf{K})-C_g(\mathbf{K})\sim (K^2/3)H^{5/2}+...
\ee

On other words, for gravity waves in shallow water at moderate shear, a single wavecrest will travel outwards in all directions at close to isotropic velocity, remaining steady and carrying its own energy outwards. Also the difference between group and phase velocities is isotropic to leading order, and so is the appearence of the resulting wave pattern. 
This is the situation shown in the bottom 8 panels of Fig.~\ref{fig_moderate}.

\subsubsection{Shallow water, strong shear}

The above conclusions change somewhat when we allow the shear to grow strong, so that $H\ll 1$ and $\Frs\gg 1$, while assuming $H\Frs\sim 1$. Now, again ignoring directions close to normal to the currents by assuming $|\cos\theta|\gg H, \Frs^{-1}$, we obtain from \eqref{cdimless}
\bs\label{dispgravshal}
\be
  C(\mathbf{K})\sim \frac{\Frs H}{2}(|\cos\theta_k|-\cos\theta)+\frac{1}{\Frs |\cos\theta|}+...
\ee
giving once again $C_g\approx C$. To wit,
\be 
  C(\mathbf{K})-C_g(\mathbf{K}) \sim K^2 H^2 \Frs\cos\theta_k /3+...
\ee
\es

We have now a phase velocity of similar appearance as that which we found for high shear in deep water, where the wave crests travel quickly along with the shear and slowly against the shear. There is a major difference in the wave pattern observed, however, because unlike that case, where group velocity was slow in all directions, here the group velocity is similar to the phase velocity. Similar to the previous shallow water case, most of the wave potential energy again resides in a single oval shaped wave crest which remains steady in shape for a long time. The picture is similar to the deep water situation in Fig.~\ref{fig_gravDeep} in directions against the current (towards the right in the figure), but strikingly different in the downstream directions (towards the left). 

\begin{figure}[htb]
  \includegraphics[width=\textwidth]{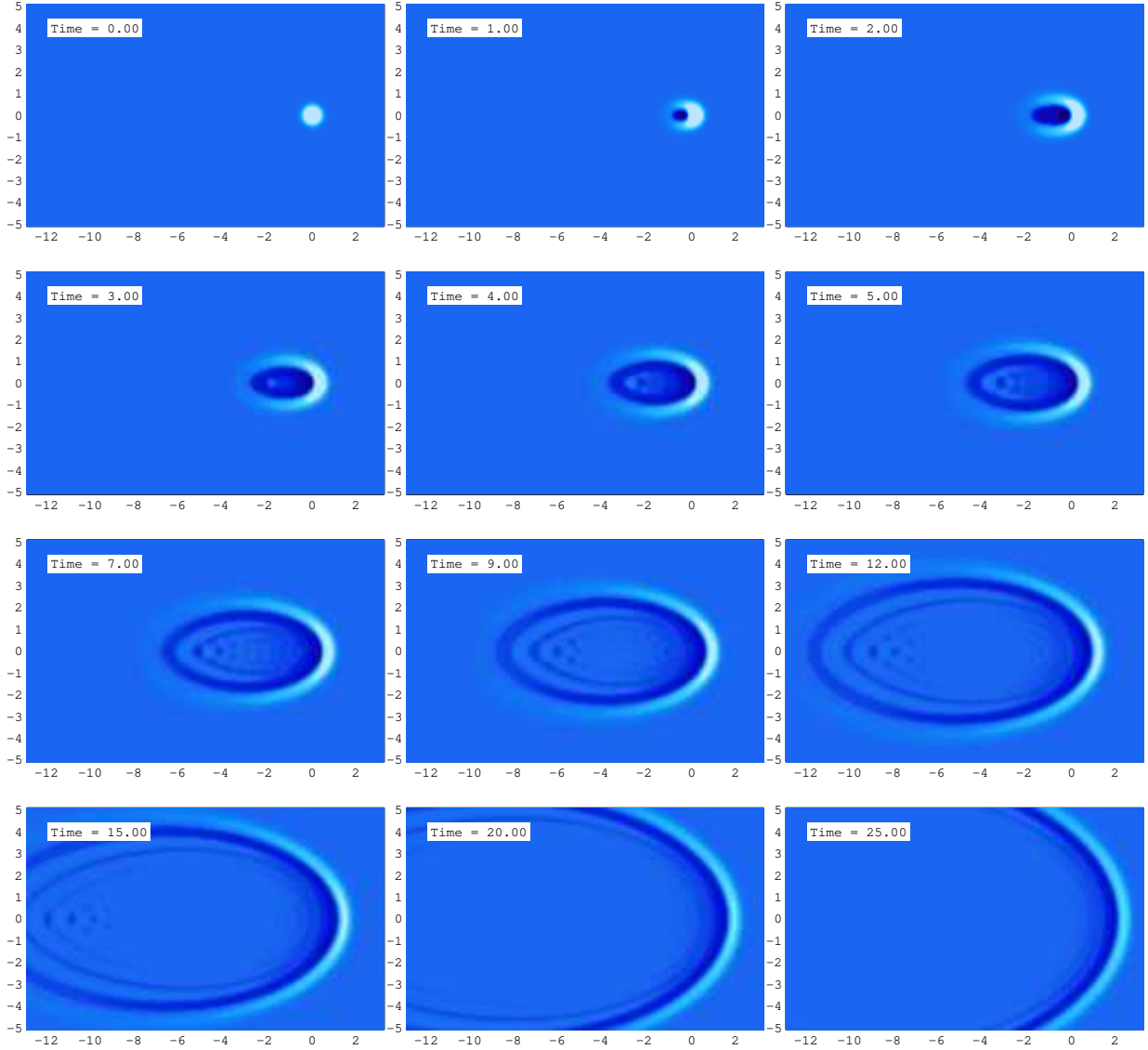}
  \caption{Relief plots of gravity waves with strong shear in shallow waters for different values of dimensionless time $T$, parameters $\Frs=10$, $\Bd=\infty$, $H=0.1$. Colour shading as in Fig.~\ref{fig_moderate}.(Multimedia view)}
  \label{fig_gravShal}
\end{figure}

The visible differences between the wave patterns in Figs.~\ref{fig_gravDeep} and \ref{fig_gravShal} all occur in propagation directions $\cos\theta_k<0$, that is, waves travelling towards the left in the figures. The reason is clear to see from dispersion relations \eqref{dispgravdeep} and \eqref{dispgravshal}: whenever $\cos\theta_k>0$, the phase and group velocities are depth independent to leading order in $\Frs^{-1}$ and therefore identical whether the waters are deep or shallow. Hence also the wave pattern will be virtually identical for the wave crests travelling against the shear current, but very different for waves travlling with the flow. Graph B in Fig.~\ref{fig_cdif} shows that the difference between phase and group velocities is far smaller than for deep water, resulting in relatively stable oval ring-shapes.

\subsection{Gravity--Capillary waves}

The picture changes once again if in addition to strong shear we assume the waves be affected by the surface tension force. Looking at the dispersion relation \eqref{cdimless} we note that the term containing $\Bd$ becomes similar to its corresponding gravity term if $\Bd\sim K^2$, so in this section we choose $\Bd = (2\pi)^2$ so that gravity and surface tension both contribute approximately equally as wave driving forces. 

\begin{figure}[htb]
  \includegraphics[width=\textwidth]{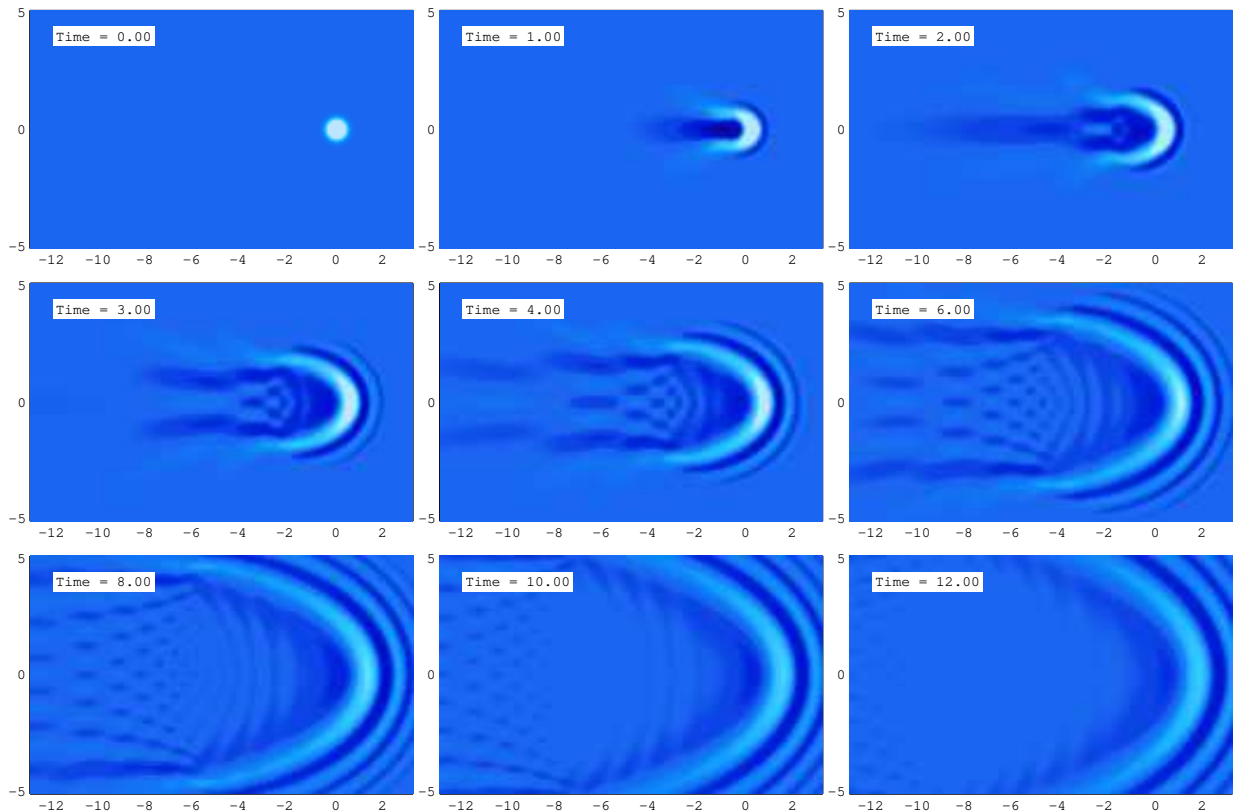}
  \caption{Relief plots of capillary--gravity waves with strong shear in deep waters for different values of dimensionless time $T$, parameters $\Frs=10$, $\Bd=4\pi^2$, $H=10$. Colour shading as in Fig.~\ref{fig_moderate}. (Multimedia view)}
  \label{fig_gravcapDeep}
\end{figure}

\begin{figure}[htb]
  \includegraphics[width=\textwidth]{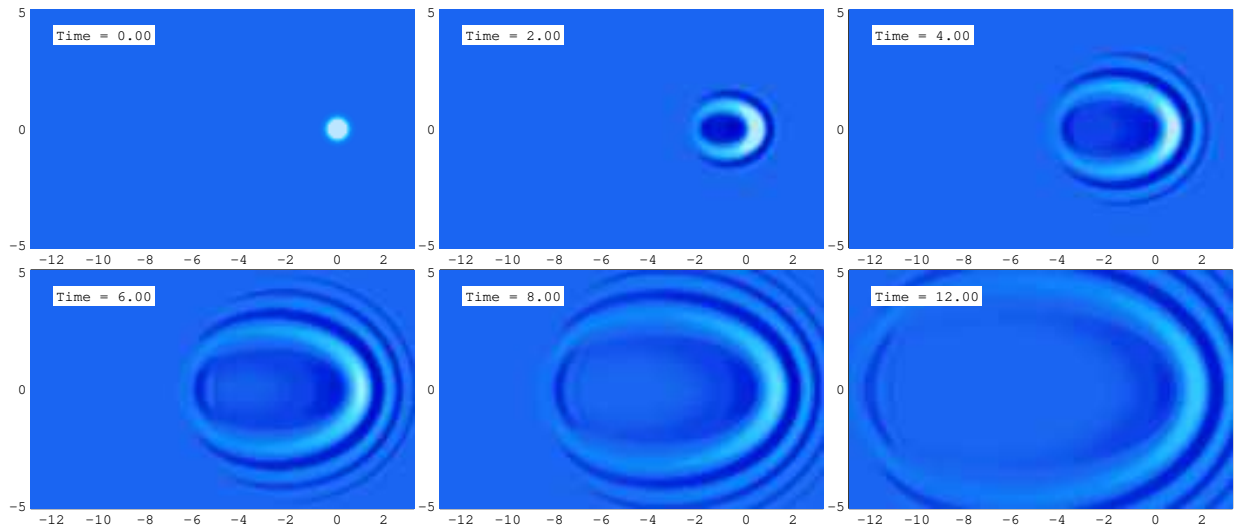}
  \caption{Capillary--gravity waves with strong shear in shallow waters for different values of dimensionless time $T$, parameters $\Frs=10$, $\Bd=4\pi^2$, $H=0.1$. Colour shading as in Fig.~\ref{fig_moderate}. (Multimedia view)}
  \label{fig_gravcapShallow}
\end{figure}

The situation is shown in figure \ref{fig_gravcapDeep} (deep water $H=10$) and \ref{fig_gravcapShallow} (shallow water $H=0.1$). 
%
%
Comparing with the pure gravity situation with the same $\Frs$ shown in figure \ref{fig_gravDeep}, we can observe the effects that the capillary force has. Most notable is that the waves moving against the shear no longer form a single long-lived crest, but that ripples travel ahead of the main crest towards the right in the panels of Figs.~\ref{fig_gravcapDeep} and \ref{fig_gravcapShallow} at a group velocity which exceeds the phase velocity; see also graph C in Fig.~\ref{fig_cdif}.

Looking again at the dispersion relation when $\Frs\gg 1$ and $H\gg 1$, assuming $K\sim 2\pi$ and $|\cos\theta_k|\gg H^{-1},\Frs^{-1}$,
\begin{subequations}\label{gravcap}
\begin{align}
  C(\mathbf{K}) =& \frac{\Frs}{2K}(|\cos\theta_k|-\cos\theta_k) + \frac{1+K^2/\Bd}{\Frs|\cos\theta_k|} + ... \\
  C_g(\mathbf{K}) =& \frac{1+3K^2/\Bd}{\Frs|\cos\theta_k|} +...
\end{align}
\end{subequations}
As for the gravity waves, the phase velocity is significantly higher along the flow than against the flow, while the group velocity is about the same in both these directions. We see that in directions travelling towards the right in Fig.~\ref{fig_gravcapDeep}, $C_g$ is about twice as large as $C$ when $K^2\approx \Bd$. This fact is responsible for the different pattern observed compared to the pure gravity waves in Fig.~\ref{fig_gravDeep}, where $C_g\approx C$ for right-propagating waves. Also the interference patterns formed by downstream (left) propagating waves are different in Figs.~\ref{fig_gravDeep} and \ref{fig_gravcapDeep}. Graph C in Fig.~\ref{fig_cdif} shows that we have ``capillary--type'' dispersion with $c_g>c$ against the shear flow, but ``gravity--type'' dispersion with $c_g<c$ in downstream directions. 

For gravity--capillary waves in \emph{shallow} water with strong shear ($\Frs\gg 1$ and $H\ll 1$) assuming $\Frs H\sim 1$, the dispersion relation is instead
\begin{subequations}
\begin{align}
  C(\mathbf{K}) =& \frac{\Frs H}{2}(|\cos\theta_k|-\cos\theta_k) + \frac{1+K^2/\Bd}{\Frs|\cos\theta_k|} + ... \\
  C_g(\mathbf{K}) =& \frac{\Frs H}{2}(|\cos\theta_k|-\cos\theta_k) +\frac{1+3K^2/\Bd}{\Frs|\cos\theta_k|} +...
\end{align}
\end{subequations}
The key difference from \eqref{gravcap} is that now phase and group velocities are similar in downstream directions (towards the left), hence a pattern of fairly stable oval shaped crests results even though the shear is great. This is shown in figure \ref{fig_gravcapShallow}, and also curve D in Fig.~\ref{fig_cdif}. Even with a high shear Froude number ($\Frs=10$ in figure \ref{fig_gravcapShallow}) the outline of the ring pattern as a whole is close to circular, although the centre of the near-circle moves down stream. This may be deduced from noting that
\[
  C_g(\bK)\approx \sqrt{H(1+3K^2/\Bd)+(\frac12 \Frs H\cos\theta_k)^2}-\frac12 \Frs H\cos\theta_k.
\]
The arguments in section \ref{sec_circs} imply that this group velocity describes a wave pattern which is contained within a circle whose radius expands at velocity $\sqrt{H(1+3K^2/\Bd)}$, and whose centre moves downstream at velocity $\Frs H/2$.

Clearly surface tension and shallow water conspire to reduce the effects of the shear on the wave appearence in the downstream direction. Note again how the dispersion relation is identical to leading order for deep and shallow waters in propagation directions against the shear currents ($\cos\theta_k>0$), giving almost identical patterns for left-propagating waves in Figs.~\ref{fig_gravcapDeep} and \ref{fig_gravcapShallow}.

\subsection{Capillary waves}

We finally consider the case where capillary forces dominate gravitational forces, quantified by a small value of the Bond number, $\Bd\ll 1$.  Consider first the case of deep water, $H\gg 1$ with moderate shear $\Frs\lesssim 1$, and again consider waves of wavelength similar to the initial disturbance, $K\sim 2\pi$, which contain the bulk of the energy. In this limit,
\begin{subequations}
\begin{align}
  C(\mathbf{K})\sim& \begin{cases}\sqrt{K/\Bd}\Bigl[1-\frac12\Frs\bigl(\Bd/K^3\bigr)^{1/2}\cos\theta_k\Bigr]+..., &H\gg 1 \\
  K\sqrt{H/\Bd}\Bigl[1 -\frac12\Frs \sqrt{\Bd H}\cos\theta_k\Bigr]+..., &H\ll 1
  \end{cases},\\
  C_g(\mathbf{K})\sim& \begin{cases}\frac{3}{2}\sqrt{K/\Bd} + ...,  &H\gg 1 \\
  2K\sqrt{H/\Bd} -\Frs H\cos\theta_k/2+..., &H\ll 1
  \end{cases}.
\end{align}
\end{subequations}
where the first correction term is of order $\sqrt{\Bd}$. 

Whether the water is deep or shallow or something in between, the shear Froude number always appears only in the sub-leading term of $C(\bK)$ with respect to $\Bd$. For waves driven primarily by capillary forces, thus, the presence of the shear is hardly felt, except for a slight directional dependence of the phase velocity at sub-leading order, moving the pattern downstream as detailed in section \ref{sec_circs}. We note that both for deep and shallow waters, the expansion of $C$ in orders of $\Bd$ has the form $C=C_\text{leading}(K)(1-\epsilon\cos\theta_k)+...$ with $\epsilon\ll 1$. In other words, the wave crests of wave number $K$ move downstream at a small velocity $\epsilon C_\text{leading}(K)$.

The fact that capillary waves feel the presence of a shear current less than do gravity waves agrees with observations made previously, e.g.\ in Ref.~\onlinecite{ellingsen13a}. 
Physically, the lack of coupling between capillary waves and shear can be understood by noting that capillary waves are short in wavelength, not much longer than the capillary length $\sqrt{\sigma/\rho g}$, about $2.7$mm for a water surface. The wave amplitude is by assumption significantly smaller than this in linear wave theory, so capillary waves hardly penetrate into the body of the water. In the thin surface layer where the capillary waves are felt, the shear current velocities are much smaller than the wave's own velocity $c$, and can only perturb the waves slightly.

\subsubsection{Deep water, strong shear}

\begin{figure}[htb]
  \includegraphics[width=\textwidth]{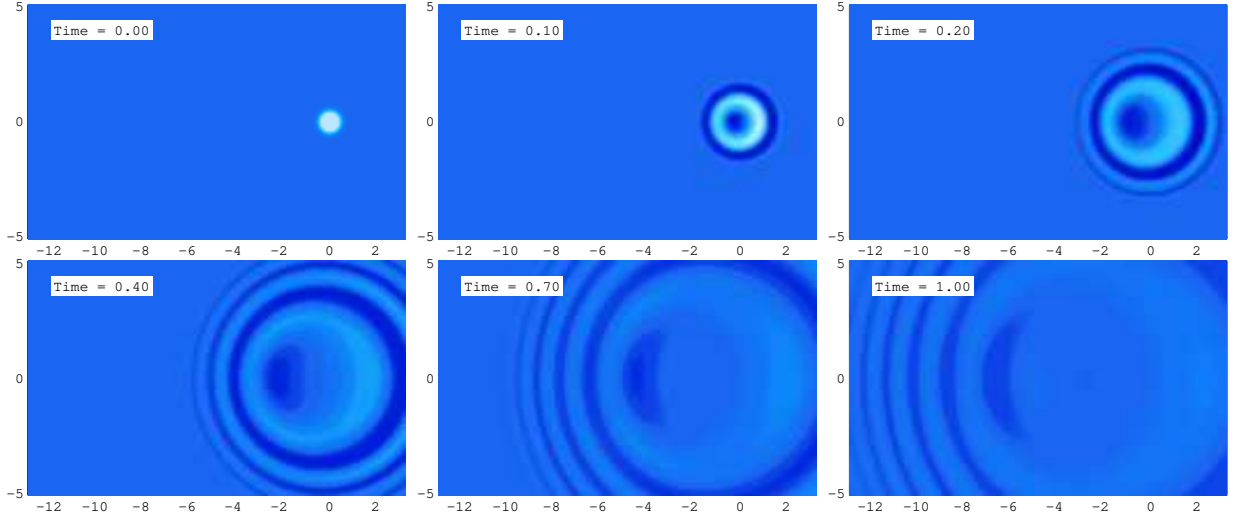}
  \caption{Relief plots of capillary waves with strong shear in deep waters for different values of dimensionless time $T$, parameters $\Frs=10$, $\Bd=0.1$, $H=10$. Colour shading as in Fig.~\ref{fig_moderate}. (Multimedia view)}
  \label{fig_capDeep}
\end{figure}

Only when the shear Froude number is increased so that $\Bd\,\Frs\sim 1$, does the shear current influence the qualitative aspects of the ring wave pattern perceptibly. Again considering deep waters ($H\gg 1$),
\bs
\begin{align}
  C(\mathbf{K})\sim& \sqrt{\frac K\Bd + \Bigl(\frac{\Frs\cos\theta_k}{2K}\Bigr)^2}-\frac{\Frs\cos\theta_k}{2K},\label{Ccds}\\
  C_g(\mathbf{K})\sim& \frac{3K^2/\Bd}{\sqrt{K^3/\Bd+(\frac12 \Frs\cos\theta_k)^2}} .
\end{align}
\es
We may now not expand the square roots, since neither term is sure to dominate the other. Both phase and group velocities are now reasonably isotropic (see graph E in Fig.~\ref{fig_cdif}) and the overall wave pattern is circular in appearance, as seen in Fig.~\ref{fig_capDeep}.

As is characteristic of capillary waves, the group velocity exceeds the phase velocity, and the waves of wavelength $\sim 1$ move significantly faster than do correspondong gravity waves (measured in dimensionless time $T$). If we interpret the pattern as a series of phase-rings emanating from a moving centre, such as we did in section \ref{sec_circs}, we find that the centre of the ring pattern moves downstream at velocity 
\[
  C_\text{centre}\sim \sqrt{\Frs \Bd}\sqrt{\frac{\Frs}{4K^3}}
\]
while the radius of each such phase-circle increases at velocity
\[
  C_\text{radius} \sim \sqrt{\frac{K}{\Bd}}.
\]
For $K\sim 2\pi$ and $\Frs\sim\Bd^{-1}\sim 10$ (the example chosen for Fig.~\ref{fig_capDeep}), we then have $C_\text{radius}\gg C_\text{centre}$ ($C_\text{radius}\approx 7.9$ while $C_\text{centre}\approx 0.10$ for the parameters of Fig.~\ref{fig_capDeep}), so each circle is seen to expand much more quickly than it moves downstream. A dispersive effect is also visible here, namely how the longer wavelength ``phase circles'' move downstream more quickly than the shorter-wavelength circles, creating the asymmetry within the pattern observed in Fig.~\ref{fig_capDeep}.

\subsubsection{Shallow water, strong shear}

\begin{figure}[htb]
  \includegraphics[width=\textwidth]{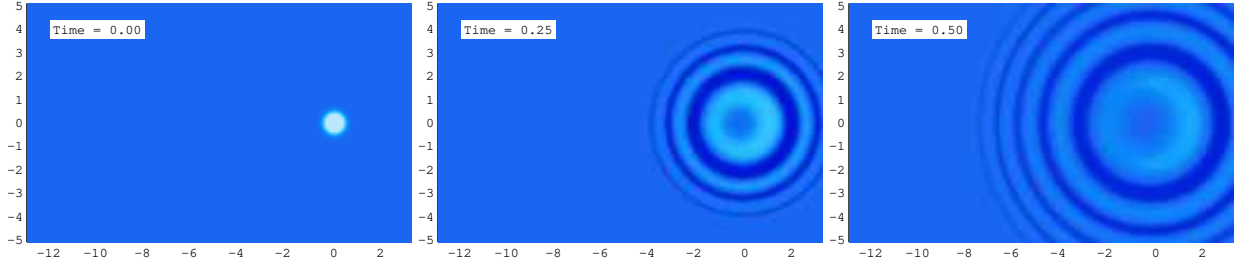}
  \caption{Relief plots of capillary waves with strong shear in shallow waters for different values of dimensionless time $T$, parameters $\Frs=10$, $\Bd=0.1$, $H=0.1$. The effect of the shear is modest. Colour shading as in Fig.~\ref{fig_moderate}.}
  \label{fig_capShal}
\end{figure}

The final case we consider is capillary waves in shallow water; $\Bd\ll 1, H\ll 1, \Frs\gg 1$. If we assume $\Bd\Frs\sim H \Frs\sim H/\Bd\sim1$, then
\bs
\begin{align}
  C(\mathbf{K})\sim& \sqrt{\frac{H}{\Bd}K^2+\left(\frac{\Frs H}{2}\cos\theta_k\right)^2}-\frac{\Frs H}{2}\cos\theta_k+...\\
  C_g(\mathbf{K})\sim& C(\bK)+ \frac{H}{\Bd}\frac{K^2}{\sqrt{(H/\Bd)K^2+(\Frs H\cos\theta_k/2)^2}}+...
\end{align}
\es
Here all large parameters appear multiplied with a correspondingly small parameter, and no extraordinary behavior can be ovserved even for extreme values. The resulting wave pattern is influenced by the presence of the shear current, but not dramatically so, as shown in figure \ref{fig_capShal}. The group velocity exceeds the phase velocity by a factor close to $2$ in all propagation directions, as seen in graph F in Fig.~\ref{fig_cdif}.

\section{Conclusions}

We have solved the Cauchy--Poisson problem of an initial disturbance of a fluid surface when a current with uniform vorticity is present. The solution was obtained by solving the full Euler equation to linear order in the wave perturbation. In particular, the dispersion relation for a plane wave propagating in an arbitrary direction relative to the current is derived. We showed that wave crests of a particular wavelengths propagate outwards in a circle whose radius expands as though no shear were present, but whose centre moves with the shear current. The dispersion properties, the group velocity in particular, are however profoundly affected by the presence of the shear, resulting in highly asymmetrical and non-circular wave patterns in some parameter regimes.

Numerical calculations were performed for the case of a Gaussian bell-shaped initial disturbance, and this case was analysed for values of the three non-dimensional system parameters, which are the (non-dimensionalised) water depth, a ``shear Froude number'' and the Bond number describing the relative influence of surface tension. In each case asymptotic analysis of phase and group velocities were employed to explain the patterns found. 

The influence of the shear current is most dramatic for gravity waves at deep water, where the upstream and downstream wave properties can differ vastly. A single long-lived crest moves very slowly upstream, while the wave train appears highly volatile in the downsream direction. Both shallower water and the influence of capillary forces were found to dampen the qualitative influence of the shear current on the pattern, with the opposite extreme being capillary-dominated shallow-water waves, whose qualitative behavour were found to be largely unaffected by even a strong shear current.

\subsection*{Acknowledgements}

The author has benefited from discussions with Professor Peder Tyvand.

\end{document}